\documentclass[aps,twocolumn,showlabels,showrefs,amsmath,amssymb,prl,superscriptaddress,floatfix,colors]{revtex4-2}

\usepackage{lineno}
\usepackage{graphicx}
\usepackage{dcolumn}
\usepackage{bm}
\usepackage{cancel}
\usepackage{graphicx}

\usepackage{dcolumn}
\usepackage{bm}
\usepackage{amssymb}
\usepackage[dvipsnames]{xcolor}
\usepackage[colorlinks]{hyperref}
\hypersetup{citecolor=Blue, urlcolor=Blue, linkcolor=Blue}

\usepackage{multirow}
\usepackage{color}
\usepackage[normalem]{ulem}

\usepackage[cp1251]{inputenc}
\usepackage{comment}

\makeatletter
\newcommand*{\sumcirclearrowleft}{%
 \DOTSB
 \mathop{
  \mathchoice
   {\rlap{\kern.25em\rotatebox[origin=c]{-90}{$\circlearrowleft$}}{\sum}}
   {\vcenter{\rlap{\kern.2em\rotatebox[origin=c]{-90}{$\scriptscriptstyle\circlearrowleft$}}}{\sum}}
   {\sum}{\sum}
 }\slimits@
}

\newcommand*{\sumcirclearrowright}{%
 \DOTSB
 \mathop{
  \mathchoice
   {\rlap{\kern.25em\rotatebox[origin=c]{90}{$\circlearrowright$}}{\sum}}
   {\vcenter{\rlap{\kern.2em\rotatebox[origin=c]{90}{$\scriptscriptstyle\circlearrowright$}}}{\sum}}
   {\sum}{\sum}
 }\slimits@
}
\makeatother

\begin{document}
\title{Depinning and activated motion of chiral self-propelled robots}

\author{Juan Pablo Carrillo-Mora}
    \email{jpcarrillo-mora@ub.edu}
    \affiliation{Computing and Understanding Collective Action (CUCA) Lab, Condensed Matter Physics Department, Universitat de Barcelona, Mart\'i i Franqu\`es 1, E08028 Barcelona, Spain}
    \affiliation{University of Barcelona Institute of Complex Systems (UBICS), Mart\'i i Franqu\`es 1, E08028 Barcelona, Spain}
\author{Adri\`a Garc\'es}
    \affiliation{Computing and Understanding Collective Action (CUCA) Lab, Condensed Matter Physics Department, Universitat de Barcelona, Mart\'i i Franqu\`es 1, E08028 Barcelona, Spain}
    \affiliation{University of Barcelona Institute of Complex Systems (UBICS), Mart\'i i Franqu\`es 1, E08028 Barcelona, Spain}
\author{Demian Levis}
    \email{levis@ub.edu}
    \affiliation{Computing and Understanding Collective Action (CUCA) Lab, Condensed Matter Physics Department, Universitat de Barcelona, Mart\'i i Franqu\`es 1, E08028 Barcelona, Spain}
    \affiliation{University of Barcelona Institute of Complex Systems (UBICS), Mart\'i i Franqu\`es 1, E08028 Barcelona, Spain}

\begin{abstract}

We study experimentally, numerically and analytically, the dynamics of a chiral active particle (cm-sized robots), pulled at a constant translational velocity. 
We show that the system can be mapped to a Brownian particle driven across a periodic potential landscape, and thus exhibits a rotational depinning transition in the noiseless limit, giving rise to a creep regime in the presence of rotational diffusion. 
We show that a simple model of chiral, self-aligning, active particles accurately describes such dynamics. The steady-state distribution and escape times from local potential barriers, corresponding to long-lived orientations of the particles,  can be computed exactly within the model and is in excellent agreement with both experiments and particle-based simulations, with no fitting parameters. 
Our work thus consolidates such self-propelled robots as a model system for the study of chiral active matter, and highlights the interesting dynamics arising from the interplay between external and internal driving forces in the presence of a self-aligning torque. 

\end{abstract}

\maketitle

\begin{figure}
    \centering
    \includegraphics[width=0.92\linewidth]{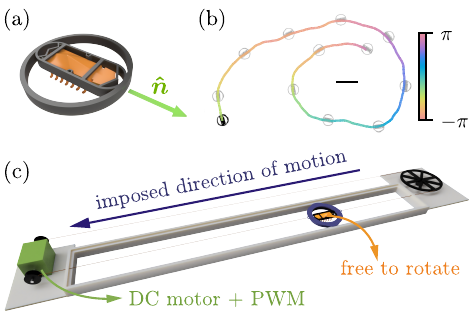}
    \caption{(a) A self-propelled robot with a director $\hat{\textbf{n}}$ in a circular shell adapted to its shape. (b) Single robot trajectory, with a color scale indicating its  orientation. The scale bar has a length of 10 cm. (c) Schematic representation of the experiment illustrating how we impose 
    a constant velocity - in a channel inside which the robot can rotate -  controlled by a motorized system and pulse width modulation (PWM).} 
    \label{fig:sketch}
\end{figure}

In the last two decades, we have witnessed a growing interest in constructing a quantitative framework to characterize, and eventually predict, the behavior of non-equilibrium systems composed of self-driven units \cite{Marchetti2013, O2022}. 
Experiments on active colloids and grains have contributed significantly to the development of the field \cite{Poon2013, Bechinger2016, Baconnier2025}, providing a playground to explore (somehow) simple model systems. 
A popular minimalist approach is to think of those as Active Brownian Particles (ABP), performing a persistent random walk \cite{Romanczuk2012}, on top of which, other physical ingredients can be included. For instance, when the left-right symmetry of the particles is broken, their  motion acquires a net chirality \cite{Liebchen2022}. Examples of Chiral Active Brownian Particlas (CAP) include bacteria swimming close to surfaces \cite{Lauga2006}, L-shaped colloidal swimmers \cite{Kummel2013} or motile robots with non-uniform mass distribution \cite{Barona2024}. Of particular relevance when dealing with self-propelled grains, is self-alignment, pushing the intrinsic orientation of the particle to match the direction of its velocity \cite{Baconnier2025}. The latter is responsible for the emergence of interesting collective dynamics  \cite{deblais2018boundaries, Boudet2021,  Baconnier2022, Baconnier2023, Baconnier2024, Lazzari2024, Gu2025, Deseigne2010, Szabo2006, Malinverno2017, Lam2015}.  
In addition, self-alignment may induce  interesting dynamics  also at the individual level, such as, 
 an orbital dynamic state  within a harmonic trap \cite{Dauchot2019} or the alignment/anti-alignment with an external potential imposed by a tilted plane \cite{Ben2023}.
Self-aligning ABPs can, for instance, be experimentally realized by cm-size self-propelled toy robots called \textit{Hexbugs}{\textregistered} \cite{Hexbugs} (see Fig.~\ref{fig:sketch}(a)). They self-align as a consequence of their mass distribution and the shape of their legs, and self-propel thanks to the vibration induced by an internal motor 
 \cite{Giomi2013, deblais2018boundaries, Dauchot2019, Baconnier2022}.  These robots have been used as a model to study  different non-equilibrium phenomena, such as clustering \cite{Levay2024}, traffic jams \cite{Barois2019}, clogging \cite{Patterson2017}, magnetotaxis \cite{Sepulveda2021}, resetting \cite{Altshuler2024}, sorting \cite{Barois2020, Li2022}, synchronization \cite{Xia2024, Zheng2023, Xu2024} and self-assembly \cite{Xi2024, Boudet2021, Sanchez2018, Giomi2013, Obreque2024} of active particles.

Here, we experimentally study the interplay between chirality and self-alignment using the before-mentioned toy (Hexbugs) robots, showing that it leads to a depinning transition with an activated creep regime.
Depinning  occurs in a variety of systems where particles or extended objects are driven through environments that resist motion 
 \cite{Chauve2000}. Although this non-equilibrium transition is typically associated with disordered landscapes,  it also occurs in periodic ones \cite{Reichhardt2016, Purrello2020, Kolton2020, Ferrero2021}, such as in superionic conductors, Josephson junctions, and phase-locked loops \cite{Risken1996}. 
 Recently, depinning has also been studied in active  systems, contributing to the understanding of, for example, the emergence of spatial organization in biofilm growth, the transport properties of active  particles with Rayleigh-Helmholtz friction, the pinning of dislocations in colloidal crystals, and the dynamic phases of active particles in porous media \cite{Straube2024, Young2023, Su2023, Vansaders2020, Sandor2017, Reichhardt2018}.
We also analyze a simple model of active particles and solve exactly its  Fokker-Planck equation in steady conditions, as well as the escape times of the activated dynamics. Despite its simplicity, the model quantitatively reproduces  the experimental data, then showing that self-propelled robots indeed constitute an experimental model system to study chiral active matter.

 \begin{figure*}[ht]
    \centering
    \includegraphics{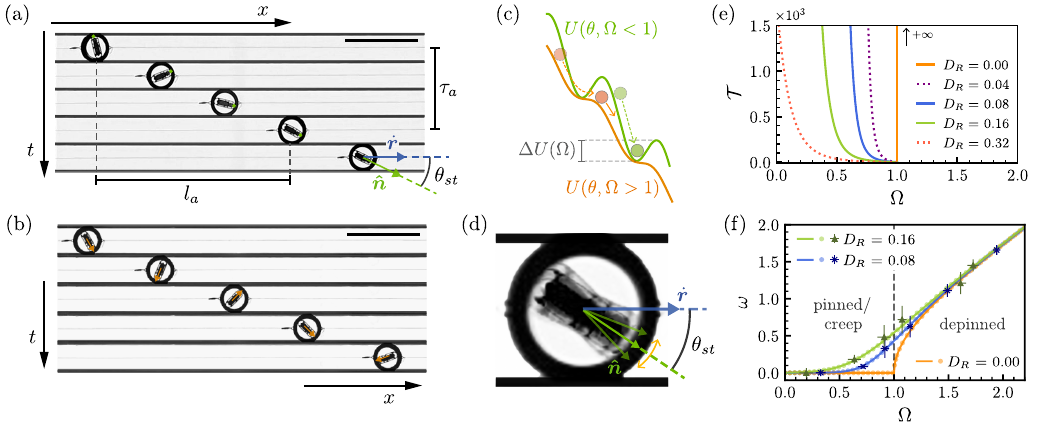}
    \caption{ 
    Experimental snapshots at different times showing (a) the relaxation of a robot's orientation towards a stationary angle $\theta_{st}$ after traveling a distance $l_a$ in a typical time $\tau_a$, when the imposed velocity is such that $|\Omega| < 1$;  (b) the rotation of the robot when the imposed velocity is small enough, i.e. $|\Omega| > 1$. 
    The scale bar in the upper right corner corresponds to 15 cm. (c) Sketch of a particle on a periodic tilted potential. For $|\Omega| < 1$ the system is pinned at the minima of the potential. For $|\Omega| > 1$ there is no longer potential minima, and the particle depins. The dynamics exhibits oscillations at a finite angular velocity $\omega$. (d) Experimental screenshot showing the fluctuations of the particle orientation around $\theta_{st}$ due to angular noise for $D_R>0$. (e) Mean first passage time $\mathcal{T}(\Omega, D_R)$ to escape the potential wells in panel (c) for $|\Omega| < 1$. (f) Dynamic phase diagram in terms of the mean angular velocity of a particle as a function of $\Omega$ for different noise intensities, showing a depinning transition with a creep regime. Continuous lines correspond to analytic results using the stationary distribution, dots to particle-based simulations and symbols with error bars to experimental measurements.}
    \label{fig:activationbarrier}
\end{figure*}

We model a self-propelled robot at position $\bm{r}$, moving on the plane, as a self-aligning chiral active particle (SA-CAP) defined by the overdamped dynamics
\begin{align}
  &  \dot{\bm{r}} = v_0\bm{\hat{n}} ,\ ~\bm{\hat{n}} = (\cos{\theta}, \sin{\theta}), \label{eq: EoM 1} \\
  &   \dot{\theta} = \omega_0 + \dfrac{|\dot{\bm{r}}|}{l_a}  \sin{\left( \phi - \theta \right)} + \sqrt{2D_R}\eta, 
    \label{eq: EoM 2}
\end{align}
where $v_0$ is the self-propulsion speed, $\bm{\hat{n}}$ is the tail-to-head director, parametrised by a phase $\theta$ with respect to an arbitrary $x$-axis, $D_R$ is the rotational diffusion coefficient and $\eta$ is a Gaussian white noise of unit variance and zero mean. 
 Chirality is introduced by $\omega_0$, the intrinsic angular velocity of the robot's orientation. The second term in Eq.~(\ref{eq: EoM 2}) stands for self-alignment. It aims to minimise the difference between the particle's orientation and its instantaneous velocity,  $\dot{\bm{r}} = |\dot{\bm{r}}|(\cos\phi,\sin\phi)$,  with a strength given by its own velocity over the length scale $l_a$, thus defining the self-alignment frequency $\omega_a=|\dot{\bm{r}}|/l_a$.

A free SA-CAP   reduces to  a Chiral Active  Particle (CAP)  as $\phi(t) = \theta(t)$, $\forall t$ \cite{Van2008, Liebchen2017}. 
Experimentally, one can track the position and orientation of a single robot (Fig.~\ref{fig:sketch}(b)), to measure $v_0$, $\omega_0$ and $D_R$ as follows: (i) From the mean squared displacement at short-times, $\langle\bm{r}(t_0)-\bm{r}(t+t_0) \rangle^2 \approx v_0^2t^2$, we extract $v_0$; (ii)  from the orientation autocorrelation function
 $\langle \bm{\hat{n}}(t) \cdot \bm{\hat{n}}(t_0) \rangle = \langle \cos{\left[ \theta(t) - \theta(t_0) \right]}\rangle=e^{-D_R t}\cos{(\omega_0 t)}$, 
 we measure $\omega_0$ and $D_R$ (see Supplemental Material \cite{SM}, Sec.~I, for details and data). 
The self-aligning characteristic length $l_a$, cannot be simply extracted from free trajectories. 
 To measure it, we drive the system externally. 
We design an experimental set-up  which imposes on a particle a given constant velocity $\bm{V}=V\bm{\hat{e}}_x$, defining the $x$-direction,  while allowing it to rotate freely (see Fig.~\ref{fig:sketch}(c)). The robot is attached to a circular shell, inside which it can freely rotate. It is then confined to a straight channel of width comparable to the diameter of the circular shell, imposing a one-dimensional motion. A wire is attached to the  robot's shell, and pulled on the other end by a DC motor, whose rotating frequency controls the velocity $V$ at which the particle is driven.  In this setup, the translational dynamics is simply given by $\dot{\bm{r}} = V\bm{\hat{e}}_x$, while the angular dynamics reads
\begin{equation}
    \dot{\theta} = \omega_0 - \omega_a \sin{\theta} + \sqrt{2D_R}\eta,
    \label{eq: rotational EoM with translational forcing}
\end{equation}
with $\omega_a = V/l_a=1/\tau_a$.
One can solve Eq.~(\ref{eq: rotational EoM with translational forcing}) in the noise-less limit ($D_R=0$) and  obtain, for $|\omega_0| < \omega_a$:
\begin{equation}
    \tan{\dfrac{\theta}{2}} = \dfrac{\omega_a}{\omega_0} + \sqrt{\dfrac{\omega_a^2}{\omega_0^2}-1} \coth{\left(-\frac{\omega_at+C}{2}\sqrt{1-\dfrac{\omega_0^2}{\omega_a^2}}\right)},
    \label{eq: chiral relaxation}
\end{equation}
where $C$ is an integration constant fixed by the initial condition. Then, having fixed $\omega_0$ from the robots' free trajectories, one can use Eq.~(\ref{eq: chiral relaxation}) to fit the experimental time series of $\theta(t)$ under forcing at different values of $V$ to extract $\omega_a$ as a function of $V$, and then $l_a = V/\omega_a$ (see SM \cite{SM} for details). Following this procedure, we measure the parameters of the model for 20 Hexbugs. They are quite heterogeneous: we measure values of $v_0$ in the range $[6.2, 11.7]$ cm/s, $D_R \in [0.01, 0.16]$ s$^{-1}$, $\omega_0 \in [-1.72, 1.06]$ s$^{-1}$ and $l_a \in [5.8, 13.9]$ cm, with broad, somehow uniform distributions. The complete data set is available in \cite{SM}.

Unlike the non-chiral case, the stationary angle $\theta_{st} = \theta(t\to\infty)$ at which the polarity of a chiral particle relaxes is in general non-zero (see Fig.~\ref{fig:activationbarrier}(a)). 
Either taking the  $t\to\infty$ limit in Eq.~(\ref{eq: chiral relaxation}), 
or alternatively, from the torque balance of the deterministic part of Eq.~(\ref{eq: rotational EoM with translational forcing}), it is found that the angle at which the polarity of the chiral particle relaxes monotonically is the stable fixed point 
\begin{equation}\label{eq:Thetass}
    \theta_{st} = \arcsin{\Omega}\,, \, ~\Omega = \omega_0/\omega_a\ .
\end{equation}
Expanding  Eq.~(\ref{eq: chiral relaxation}) around $\Omega = 0$, one recovers the exponential relaxation reported in \cite{Baconnier2022} and thus $\theta_{st} \to 0$ (see  \cite{SM}).

A saddle--node bifurcation takes place at $|\Omega| = 1$ (i.e. $|\omega_0| = \omega_a$). As $|\Omega|\to 1$  the stable fixed point $\theta_{st}$ of Eq.~(\ref{eq:Thetass}) annihilates with the unstable one $\pi-\theta_{st}$, meaning that the system goes from a phase-locked to a continuously rotating state (see Fig.~\ref{fig:activationbarrier}(b)). 
In fact, when $|\Omega| > 1$, Eq.~(\ref{eq: chiral relaxation}) becomes complex and the real part of the explicit solution is periodic, with period $T = 2\pi/\sqrt{\Omega^2 - 1}$ and mean angular velocity $\langle\dot{\theta}\rangle \equiv \omega \sim \sqrt{\Omega^2 - 1}$ (see details in \cite{SM}). Expanding it close to the bifurcation point (from above, $\Omega \to 1^+$), we find
\begin{equation}
    \omega \sim (\Omega - \Omega_c)^\beta,
\end{equation}
showing the presence of a depinning transition at the critical value $\Omega_c = 1$, with critical exponent $\beta = 1/2$, as typically found in this type of non-equilibrium phenomenon \cite{Kolton2020,Ferrero2021, Straube2024} and shown in Fig.~\ref{fig:activationbarrier}(f).

Note that, one can rewrite Eq.~(\ref{eq: rotational EoM with translational forcing}) as the equation of motion of a Brownian particle in a tilted periodic potential: $\dot{\theta} = -U'(\theta) + \sqrt{2D_R}\eta$, with $U(\theta) = -\omega_0\theta -\omega_a\cos{\theta}$. 
We are thus left with a Brownian particle in a corrugated potential, driven by a constant force \cite{Purrello2020, Kolton2020}. In the present case,  both the fluctuations and the `tilt' come from the intrinsic activity delivered by the robot.  
%
%
The above mapping allows us to qualitatively explain the depinning transition observed at the zero-noise limit. As shown in Fig.~\ref{fig:activationbarrier}(c), for $|\Omega| < 1$, the potential has a periodic series of minima, corresponding to stable configurations. These are \emph{pinned} states, characterized by the steady orientations given by Eq.~(\ref{eq:Thetass}). In this $|\Omega| < 1$ regime, a robot relaxes its orientation to one of these minima, as illustrated in Fig.~\ref{fig:activationbarrier}(a). 
For $|\Omega| > 1$, the self-aligning torque is no longer able to balance the chiral one. The potential is monotonic and thus the resulting dynamics unbound: the particle slides through the potential, i.e. it rotates as shown in Fig.~\ref{fig:activationbarrier}(b). The system is \emph{depinned}. The transition between the two regimes, pinned and depinned, occurs precisely when $|\Omega| = 1$, the value for which the minima of $U$ become inflection points.

So far, rotational noise has been ignored. As well documented studies of depinning, fluctuations, typically thermal in this context, round the transition and give rise to the so-called creep regime \cite{Ferrero2021creep}. In experiments, rotational noise is, of course, present and introduces fluctuations in the robots orientation, as illustrated in Fig.~\ref{fig:activationbarrier}(d). In the effective potential picture, these correspond to fluctuations around the minima in the $|\Omega| < 1$ regime. If the noise amplitude $D_R$ is large enough, the particle might escape the potential well of height $\Delta U$. Such activation process will thus allow the particle to jump from well to well, corresponding to a 2$\pi$ rotation of the robot to go back to the steady orientation. The dynamics activated by angular noise consists of a stochastic alternation between the fluctuating state around $\theta_{st}$, a minimum of $U$, and a complete rotation moving the system to the next potential minimum at $\theta_{st}+2\pi$. 

To further quantify the dynamics, we compute exactly the escape times associated with the activation process across a potential barrier as a mean first passage time $\mathcal{T}$ \cite{Peters2017} (see SM \cite{SM} for the details of the derivation). In the $|\Omega| \ll 1$ limit, it can be estimated from Kramers' theory \cite{Hanggi1990}
\begin{align}\label{eq:mfpt-final}
    \mathcal{T} = \frac{2\pi}{\omega_a (1-\Omega^2)}\exp\left(\frac{\Delta U(\Omega)}{D_R}\right),
\end{align}
where,
    $\Delta U(\Omega) = \omega_a\left[\Omega\left(2\arcsin \Omega - \pi\right) + 2\sqrt{1-\Omega^2}\right]$
is the associated energy barrier (see Fig.~\ref{fig:activationbarrier}(c)), a decreasing function of $|\Omega|$ such that $\Delta U(|\Omega| = 1) = 0$. 
As we show in \cite{SM}, Eq.~(\ref{eq:mfpt-final}) provides a very good approximation of $\mathcal{T}$ for $|\Omega| \ll 1$. As for a Brownian particle (in contact with a thermal bath at temperature $T$)  \cite{Kramers1940, Hanggi1990}, the mean escape time over a potential barrier grows exponentially fast with the barrier height $\Delta U(\Omega)$, modulated by the amplitude of the noise. Here, $D_R$ plays the same role as $k_B T$. Now, the height of the potential barrier explicitly depends on the competition between the activity of the particle, and the strength of the drive via $\Omega$. As $0 < \Omega < 1$ increases at fixed $D_R$, the barrier height reduces, vanishing at $\Omega_c=1$, and thus also does the escape time. 

Fig.~\ref{fig:activationbarrier}(e) shows the exact $\mathcal{T}$ computed from our model as a function of $\Omega$ for different noise intensities. Of course, in the absence of noise, activation is suppressed and thus escape times $\mathcal{T}$ diverge. As $D_R$ is increased, activation events are more likely and $\mathcal{T}$ takes finite values that increase as the barrier gets higher when $\Omega$ decreases. In this creep regime, the particle can rotate even if $\Omega < \Omega_c$, and acquires a finite mean angular velocity $\omega = \langle\dot{\theta}\rangle$ in steady conditions that we represent in Fig.~\ref{fig:activationbarrier}(f) as a function of $\Omega$.  
Unlike the case where $D_R = 0$, where the depinning transition is sharp, finite noise allows for activation events that round it, leading to the creep regime \cite{Purrello2020}. The dynamic phase diagram Fig.~\ref{fig:activationbarrier}(f) includes results from experiments, simulations of Eqs.~(\ref{eq: EoM 1})--(\ref{eq: EoM 2}) and analytical calculations, all of them in good agreement. 

The analytical results Fig.~\ref{fig:activationbarrier}(f) are obtained from the steady-state solution $P_{st}(\theta)$ of the Fokker-Planck equation of our model:
\begin{equation}
    \dfrac{\partial P_{st}}{\partial t} =0= -\dfrac{\partial}{\partial \theta}\left[\left( \omega_0 - \omega_a\sin{\theta}\right)P_{st}\right] + D_R \dfrac{\partial^2P_{st}}{\partial \theta^2}\,,
    \label{eq: FP equation}
\end{equation}
that can be solved exactly (see details in \cite{SM}):
\begin{equation}
     P_{st} = \left[ A + B \sum_{n = -\infty}^{\infty} (-1)^n I_n\left(\frac{\omega_a}{D_R}\right) \dfrac{e^{\left(in - \frac{\omega_0}{D_R}\right)\theta}}{in - \frac{\omega_0}{D_R}} \right] e^{-\frac{U(\theta)}{D_R}}\,.
    \label{eq: stationary distribution}
\end{equation}
Here $I_n(x)$ is the first kind modified Bessel function of order $n$, and $\lbrace A, B \rbrace$ are constants fixed by the periodicity and normalization conditions \cite{Risken1996}. Then, an expression for $\omega$ can be obtained through $\omega=-\int_{-\pi}^{\pi}U'(\theta)P_{st}(\theta)d\theta$ \cite{Risken1996, Stratonovich1967}, and as shown in Fig.~\ref{fig:activationbarrier}(f), it matches the experiments and simulations in all regimes. 

\begin{figure}[ht]
    \centering
    \includegraphics[width=\linewidth]{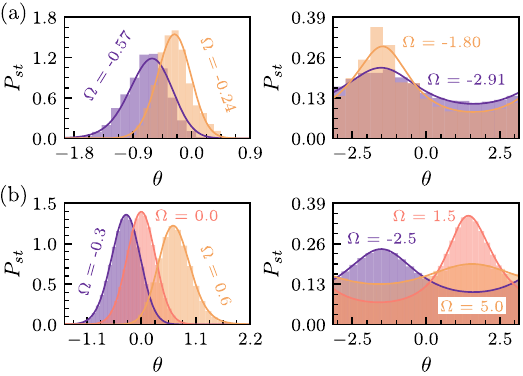}
    \caption{
    Stationary distributions for $|\Omega| < 1$ (left panels, creep regime) and for $|\Omega| > 1$ (right panels, depinned regime). 
    The histograms were obtained from  (a) experiments and (b) numerical simulations. The solid lines show the exact solution of the Fokker-Planck equation Eq.~(\ref{eq: stationary distribution}). Each distribution has a different value of $\Omega$ indicated in the figure, in all cases $D_R = 0.08$.}
    \label{fig:FP}
\end{figure}

To provide a more detailed quantitative characterization of the system,  we measure the stationary probability of finding the robot with a given orientation. The histograms in Fig.~\ref{fig:FP} show  such  distribution $P_{st}$ for different values of $\Omega$ and $D_R$,  obtained from  (a)  experiments and (b) simulations. In both cases, the analytic solution Eq.~\ref{eq: stationary distribution} reproduces the data by just setting the parameters of the model to the values previously measured in the experiments. 
In the $\Omega=0$ limit, $P_{st}$ takes the von Mises form $P_{st}(\theta) = A e^{(\omega_a/D_R)\cos{\theta}}$, the equivalent of a normal distribution on a unit circle. 
As $\Omega$ increases, the peak of the distribution, giving the most likely steady orientation, also increases, as expected from Eq.~(\ref{eq:Thetass}) when $|\Omega| < 1$. For $|\Omega| > 1$, the distribution still exhibits a maximum due to the anisotropy of the angular velocity imposed by the self-alignment torque. However, as $|\Omega|$ increases, it becomes less pronounced, and larger deviations from such maximum become increasingly likely. Indeed, in the $|\Omega| \to \infty$ (or $|\omega_0| \gg \omega_a$) limit, the distribution Eq.~(\ref{eq: stationary distribution}) becomes uniform \cite{SM}.

We have shown that a simple model of Chiral Active  Particles accurately describes the dynamics of self-propelled Hexbugs. We have measured the parameters of the model in a collection of robots,  thus allowing for a direct comparison between the experimental trajectories and the predictions of the model across a wide range of parameter values. Under a constant-velocity drive, the equations of motion of a single active particle are simple enough to allow for an analytic treatment: the steady-state distribution can be computed exactly, giving access to the dynamic phase diagram of the system. It exhibits a depinning transition, emerging from the competition between the external and internal (active torque) driving. Angular noise in the motion of the Hexbugs is responsible for a creep regime that one can, again, characterise analytically from the exact calculation of activation times in the model. Analytic predictions, particle-based simulations and experiments are in excellent agreement. Overall, our study shows that self-propelled robots constitute a good model system of chiral active matter.

\paragraph{Acknowledgements}
The authors acknowledge DURSI and Ministerio de Ciencia, Innovaci\'on y Universidades MCIU/AEI/FEDER  for financial support under Project No.~2021SGR-673 and PID2022-140407NB-C22.  A.G. acknowledges AGAUR and Generalitat de Catalunya for financial support under the call FI SDUR 2023 Ref.~CCI 2021ES05FPR011. J.P.C.M. acknowledges AEI/MICINN and FSE+ for financial support under the call FPI 2023 Ref.~PREP2022-000148 and funding from ANID Doctorado Becas Chile 2023 No.~72230400.

\bibliography{biblio}

\end{document}